\documentstyle[aps,balanced,prl]{revtex}

\begin{document}
\draft
\tightenlines

\title{Wave-unlocking transition 
 in resonantly coupled complex Ginzburg-Landau equations}
\author{A. Amengual$^*$, D. Walgraef $^*$ \cite{Daniel}, M. San
Miguel$^{*\#}$ and E. Hern\'andez-Garc\'\i a$^{*\#}$}
\address{$^*$Departament de F\'\i sica, Universitat  
de les Illes Balears, E-07071 Palma de Mallorca, SPAIN \\
$^\#$ Instituto Mediterr\'aneo de Estudios Avanzados, IMEDEA (CSIC-UIB), 
 E-07071 Palma de Mallorca, SPAIN}
\date{\today}
\maketitle
\begin{abstract}
We study the effect of spatial frequency-forcing 
on standing-wave solutions of coupled complex Ginzburg-Landau equations.  
The model considered  describes several situations of nonlinear 
counterpropagating 
waves and also of the  dynamics of polarized light waves. We show that forcing 
introduces spatial 
modulations on standing waves which remain frequency locked with a 
forcing-independent
frequency. For forcing above a threshold the 
modulated standing waves unlock, bifurcating into a temporally periodic state. 
Below the threshold the system presents a kind of excitability.  

\end{abstract}


\begin{twocolumns}

 Different physico-chemical systems driven out of equilibrium may undergo 
Hopf bifurcations leading to rich spatio-temporal behavior. When these 
bifurcations
occur with broken spatial symmetries, they induce the formation of wave 
patterns
 described by order parameters of the form :
 
 \begin{equation}
\Psi = Ae^{  i k_c x  + i \omega_c t} +Be^{ - i k_c x  + i \omega_c t} + c.c.
 \end{equation}
 
\noindent where the slow dynamics of the wave amplitudes $A$ and $B$ obey 
complex 
Ginzburg-Landau equations. This is the case, for example, for 
 Rayleigh-B\'enard convection in binary fluids, Taylor-Couette instabilities 
 between
 corotating cylinders, electro-convection in nematic liquid 
 crystals \cite{CrossHoh}, or for the transverse field of high Fresnel number 
 lasers \cite{NewMol}. 
 Symmetry breaking transitions are usually very sensitive to small 
 perturbations 
  or external fields. For example, it has been shown that a spatial 
  modulation 
  of the static electro-hydrodynamic instability of nematic liquid crystals  
  modifies the selection and stability of the resulting roll patterns.
  In particular, the constraint imposed by a periodic modulation of the 
  instability point 
   may lead to a commensurate-incommensurate phase 
transition \cite{gollub}.  In the case of Hopf bifurcations,
external fields inducing spatial or temporal 
modulations strongly affect the selection and 
stability
of the resulting spatio-temporal patterns. For example, standing waves may be 
stabilized by purely temporal modulations at twice the critical 
frequency \cite{DW,HRiecke}, or by purely spatial modulations at twice the 
critical 
wavenumber \cite{Dangel}, in regimes where they are otherwise unstable, 
including domains where the bifurcation parameter is below the critical one.

External forcings that break space or time 
translational invariance, but not the space inversion symmetry of the waves 
amplitudes, induce linear resonant couplings between the complex 
Ginzburg-Landau equations (CGLE)
which describe the dynamics of the amplitudes of left and right travelling 
waves. In the case of forcings that break the space translation invariance, 
the coupling
coefficient $\epsilon$ is in general complex, and the corresponding coupled 
CGLE may be written, in one-dimensional geometries, as :
\begin{eqnarray}
\dot A + v_g\partial_xA&=&\mu A+(1+i\alpha )\partial_x^2A \nonumber\\
      &-&(1+i\beta ) \left(\vert A\vert ^2 + \gamma \vert B\vert ^2 \right) A
      + \epsilon B\>,\nonumber\\
\dot B - v_g\partial_xB&=&\mu   B+(1+i\alpha )\partial_x^2B \nonumber\\
      &-&(1+i\beta ) \left(\vert B\vert ^2 + \gamma \vert A\vert ^2 \right) B
      + \epsilon A
\label{theEquations}
\end{eqnarray}

Due to the resonant coupling with coefficient $\epsilon$, pure travelling waves 
are not solutions any more of these equations, and
generic arguments of bifurcation theory allows a characterization of the 
possible
uniform amplitude solutions depending on the various dynamical parameters
of the system \cite{Dangel}. Here also, standing waves may be stabilized as the
result of phase locking between the waves $A$ and $B$. Predictions based on 
(\ref{theEquations}) in the $x-$independent case have been successfully tested 
for azimuthal waves in an annulus laser with imperfect O(2) symmetry 
\cite{tredicce}. However, the combined effect of the complex coupling 
coefficient
 $\epsilon$ and the spatial degrees of freedom has not been explored.

In this Letter, we study 
equations (2) with the following parameter restrictions: imaginary linear 
coupling 
coefficient  ($\epsilon =i\gamma_P$), negligible group velocity $v_g$, and weak and real 
nonlinear cross-coupling term ($\gamma < 1$). We will however maintain the 
spatial derivative in the r.h.s. of (\ref{theEquations}), and this will be 
crucial for the results below. 
We will show that the spatial forcing introduces 
spatial modulations of the standing waves solutions while $A$ and $B$ remain
frequency locked with a forcing-independent frequency.  By increasing the 
forcing, these stable 
modulated waves merge with unstable 
ones in saddle-node bifurcations with nontrivial global structure. This 
wave-unlocking
transition results in a mixed state with limit cycle temporal behavior.
The threshold value of the forcing and the limit cycle frequency are calculated
analytically. 
Modulated standing waves can also be induced by strong enough temporal
forcing \cite{Douady}.

The parameter regime explored here would be appropriate in physical 
situations where a spatial forcing modulates the frequency of the Hopf 
instability and induces a purely imaginary resonant forcing 
(a purely real $\epsilon$ would appear due to a 
spatial modulation of the distance to the instability point). Possible
systems should have negligible group velocities, as in some 
circumstances in binary fluid convection \cite{kolodner} or 
liquid crystals \cite{liqcrys}; and weak coupling such as in 
viscoelastic convection \cite{valpo}. Up to now, 
the parameter range considered here best applies to several situations
in laser physics.
A first one corresponds to taking into account transverse effects
in inhomogeneously broadened ($\gamma < 1$) bidirectional ring lasers 
\cite{singh}. The purely imaginary resonant coupling is a consequence of 
conservative (off-phase) backscattering \cite{etrich}, or alternatively, a 
spatial modulation of the refraction
index of the laser medium. In fact, a spatially periodic refractive index is 
the mechanism
used for single frequency selection in index coupled Distributed Feedback 
lasers (DFB).  A second situation is that 
of the transverse vector field in a laser near threshold \cite{maxi}. The 
parameter $\gamma_P$  corresponds to a detuning splitting between 
light linearly polarized in different orthogonal directions, produced for 
example by small cavity anisotropies. In this case, $A$ and $B$ are not the 
amplitudes of left or right travelling waves, but the amplitudes of 
the two independent circulary polarized components of light, that is 
$A=( A_x+iA_y)/\sqrt{2}$ and $B=( A_x-iA_y)/\sqrt{2}$, where $A_x$ and 
$A_y$ are the linearly polarized complex amplitudes of the vector electric 
field with a spatially transverse dependence. Weak coupling ($\gamma < 1$)
favors linear polarization 
($|A|=|B|$). We will often use the 
light-polarization terminology, because it gives a clear physical insight into 
the states found for the general set of Eqs. (\ref{theEquations}) of broad 
applicability within the parameter restrictions above.

Two families of solutions of the coupled CGLE
(\ref{theEquations})
can be distinguished.
The first family corresponds to travelling waves for $A$ and $B$ with
the same amplitude, frequency, and wavenumber:
\begin{eqnarray}
A&=& Q_0 e^{ - i k x  + i \omega t + i ( \theta_0+\psi_0 ) }\>,\nonumber\\
B&=& Q_0 e^{ - i k x  + i \omega t + i ( \theta_0 - \psi_0 ) }\>.
\label{simple}
\end{eqnarray}
Without forcing ($\gamma_P=0$), the constant global and relative phases, 
$\theta_0$ and $\psi_0$, are arbitrary, the amplitude is
$Q_0^2= (\mu - k^2)/(1+\gamma)$, and the frequency $\omega$ is 
$\omega_0 = -\alpha k^2-\beta (1+\gamma) Q_0^2$.
With forcing, the global phase and the amplitude remain unchanged, but
the relative phase is fixed by $\sin 2 \psi_0 = 0$;
the two allowed values of $\psi_0$ give two solutions 
with frequencies $\omega= \omega_0 \pm \gamma_P$.
The phase instabilities
of these solutions were discussed in \cite{maxi}. 

The second family of solutions can be searched in the form of two
waves
\begin{equation}
A = e^{i \omega_0 t} \sum_n a_n e^{inkx} \ \> , \> \ 
\ \ B = e^{i \omega_0 t} \sum_n b_n e^{inkx}\>,
\label{multiharmonic}
\end{equation}
\noindent 
frequency-locked to a frequency $\omega_0$ independent of forcing. 
For $\gamma_P=0$, the exact solutions of (\ref{theEquations}) in this form
only have two terms: $\vert a_1 \vert = \vert b_{-1} \vert = Q_0$.
The effect of a small
forcing in this solution is to generate higher harmonics, while keeping
$\omega_0$ fixed and the relative phase between $a_1$ and $b_{-1}$
arbitrary. Now, the remaining coefficients $a_n$ and $b_n$ are not zero
and can be calculated perturbatively in $\gamma_P$.
Close enough to the threshold for a mode $k$  
($\mu - k^2 \approx 0$), the amplitude of 
higher order harmonics is negligible and, to lowest 
order in $\mu-k^2$, an approximate solution takes the form

\begin{eqnarray}
 A &=& e^{i (\theta_0 + \omega_0 t)} (Q e^{i (kx + i \psi_0)} + Re^{- i (kx +
\psi_0 - \phi)} )\>, \nonumber\\
 B &=& e^{ i (\theta_0+ \omega_0 t)} ( Q e^{- i (kx + \psi_0)} + Re^{i (kx +
\psi_0 + \phi )} )\>,  
\label{mixed}
\end{eqnarray}
with $\theta_0$ and $\psi_0$ arbitrary, and $\phi$ fixed by the
forcing. $Q$ and $R$ are real numbers (positive or negative) and,
for small $\gamma_P$, $|R| \ll |Q|$ 
(an equivalent solution is found interchanging $Q$ and $R$).

A visualization of these solutions can be given within 
the polarization interpretation of (\ref{theEquations}). 
Defining $C_\pm e^{i \zeta_\pm} \equiv Q \pm e^{ i \phi} R$, the change
of variables to the amplitudes 
of the $x-$  and $y-$linearly
polarized components gives
\begin{eqnarray}
&  A _x = \sqrt{2} C_+ \cos (k x + \psi_0) e^{i ( \omega_0 t + \theta_0 +
\zeta_+)}\>, \nonumber\\
& A_y = \sqrt{2} C_- \sin (k x + \psi_0) e^{i ( \omega_0 t + \theta_0 +
\zeta_-)}\>,
\label{elipse}
\end{eqnarray} 
\null
\null

\begin{figure} 
\vskip 7.7 cm
\includegraphics{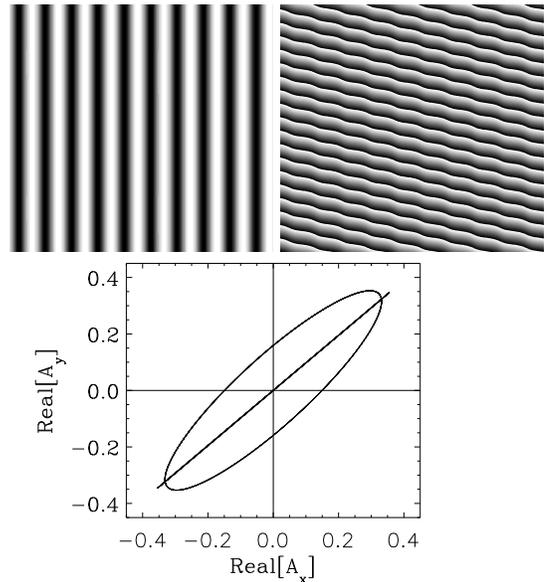}
\caption{
Top: 
Modulus (left) and phase (right) of 
$A$  for $\gamma_P=0.012 < \gamma_{Pc}$. The horizontal axis is space 
(256 units), and
the vertical is time (1000 units). Gray levels range from black (0) 
to white (the maximum
of the modulus or $2 \pi$ for the phase).
 This numerical solution 
has been obtained from (2) with
$\beta=0.2$, $\gamma=0.5$, $\mu=0.2$, $\alpha=2.6$. The 
initial condition is a standing wave with $\omega=\omega_0$ and $k=0.123$.
  Bottom: 
Polarization representation of the solution at a given point $x$. 
For $\gamma_P=0$ one 
has linear polarization (indicated by the straight line)
which becomes elliptical for $\gamma_P \not= 0$. 
An equivalent solution has the major axis of the ellipse
along the second and fourth quadrants.}
\end{figure}
\begin{figure} 
\vspace*{8.0 cm}
\includegraphics{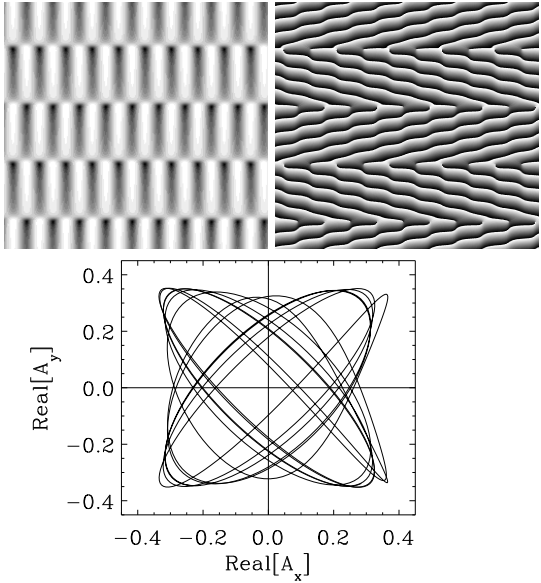}
\caption{
Same as Fig. 1 but using $\gamma_P=0.0145 > \gamma_{Pc}$.} 
\end{figure}

These equations
describe at each point $x$ the superposition of two dephased
harmonic motions with different amplitudes and a frequency $\omega_0$
independent of forcing. This identifies the solution (\ref{mixed}) with an
elliptically polarized standing wave pattern in which the orientation
of the ellipse and its ellipticity vary periodically in the spatial
coordinate $x$. In the limit of no forcing, $R = 0$, the ellipse
degenerates in a linearly polarized standing wave with an angle of 
polarization $\psi = kx + \psi_0$. In this interpretation, 
the first family of solutions, (\ref{simple}),
would correspond to linearly polarized traveling waves with frequency $\omega$
and an angle of polarization $\psi_0$. In such a case, the forcing fixes the
direction of polarization so that only $x$- or $y$-linearly polarized
waves remain. On the contrary, forcing in (\ref{mixed}) grows an ellipse 
from a linearly polarized standing wave keeping the frequency unchanged.

Elliptically polarized standing wave patterns are
obtained from a direct numerical integration of the coupled CGLE as shown 
in Fig. 1. Increasing the forcing, these solutions become unstable 
through a bifurcation in which $Q$ and $R$ become time 
dependent. As shown in Fig. 2, the solution
beyond this instability oscillates between the two equivalent
elliptically polarized standing wave patterns found for small
$\gamma_P$. In addition, from the numerical simulations, one finds 
that the period $T$ of these oscillations decreases beyond
the critical value $\gamma_{Pc}$. One has
$T^{-2} \propto (\gamma_P - \gamma_{Pc})$ (see Figs. 3 and 4).

A quantitative description of the
instability, including the determination of the critical forcing
$\gamma_{Pc}$ and the period of the oscillations, can be performed by an
amplitude analysis.
Close to the threshold for the $k$ modes, the equations 
for the slow time evolution of $Q$ and $R$ can be found by substitution 
of (\ref{mixed}) into (\ref{theEquations}) and neglecting contributions 
from higher order harmonics. Defining 
$X e^{i\Phi} \equiv  Q + i R$  we find
\begin{eqnarray}
 \dot X &=& (\mu - k^2)X -(1+\gamma)X^3 -  (1+ \gamma \cos 2\phi){X^3\over 2}
 \sin^2 2\Phi\>,\nonumber\\ 
 \dot \Phi &=&  -  (1+ \gamma \cos 2\phi){X^2\over 2}
 \sin2\Phi \cos2\Phi \nonumber\\ 
    &+&\beta\gamma  \sin 2\phi {X^2\over 2}
 \sin2\Phi -\gamma_P\sin \phi\>,\nonumber\\ 
 \dot \phi &=& \beta (1+\gamma \cos 2\phi)X^2\cos2\Phi
   +\gamma X^2 \sin 2\phi
  \nonumber\\ 
    &-&2\gamma_P\cos \phi \cot 2\Phi\>.
\label{polares}
  \end{eqnarray}
The fixed points of (\ref{polares}) represent the polarized standing 
waves solutions (\ref{mixed}). These points can be 
determined exactly in the limiting case of $\beta=0$. The interesting 
solutions have two allowed values of $\phi$: $\phi_0 = (2n +1)\pi/2$, 
$n=0$, $1$; and for each value of $\phi$, there are eight fixed points: 
Four are stable ($+$) and the other four are saddle points ($-$). 
The corresponding values of $X$ and $\Phi$ are:
\begin{eqnarray}
 X_0(\pm)^2 &=&{\mu -k^2\over 2(3+\gamma)(1+\gamma)} \nonumber\\ 
&\ & \!\!\!\!\!\!\!\!\!\! \left[
5+3\gamma \pm 
 \sqrt{(1-\gamma)^2-{8(1+\gamma)(3+\gamma)\gamma_P^2\over (\mu -k^2)^2}}
\right] \>,
\end{eqnarray}
\begin{equation}
\Phi_0(+) = \xi(+) + m\frac{\pi}{2},\  
\Phi_0(-) = \frac{\pi}{4} - \xi(-) + m\frac{\pi}{2} \ , 
\label{Phi}
\end{equation}
where $m=0,1,2,3$, and 
\begin{equation}
\xi(\pm) = 
{(-1)^{n+1}\over 4}\arcsin {4\gamma_P\over (1- \gamma )X_0(\pm)^2} \>.
\label{Phi0}
\end{equation}
Heteroclinic orbits connect the saddles and the stable nodes with the same
 $\phi$. When
$\gamma_P$ grows, saddles and 
nodes approach by pairs and at the critical value 
\begin{equation}
 \gamma_{Pc} = {(\mu-k^2)(1-\gamma)\over \sqrt{8(1+\gamma)(3+\gamma)}}\>,
\end{equation}
they merge and disappear via inverse saddle node bifurcations. 
The interesting point is the global structure of the bifurcation: the presence 
of the heteroclinic connections gives rise to the birth of limit cycles 
(one for each value of $\phi$). This is similar to the Andronov-van-der-Pol 
bifurcation \cite{Andronov} that appears in
several types of excitable systems \cite{excitable}. The difference is that, 
due to symmetries, here we have several pairs of fixed points merging, 
instead of just one pair. The periodic behaviour is illustrated in Fig. 2 
by the periodic alternation of the 
trajectory between the ``ghosts'' of the disappeared elliptically 
polarized states corresponding to the fixed points. 
Below the bifurcation, 
small perturbations around the stable solutions decay, whereas perturbations  
above a threshold push the system along the heteroclinic 
trajectory towards another stable fixed point. Since the size of the 
perturbation required for such switches decreases by increasing $\gamma_P$, 
and vanishes at $\gamma_{Pc}$, the multistability of this system can be 
seen as a kind of excitability \cite{meron}.  A different 
consequence of the multiplicity of stable states is their possible 
coexistence in
space, leading to the formation of domains with different polarizations along 
the $x$ axis.

\begin{figure} 
\vskip 4.2 cm
\includegraphics{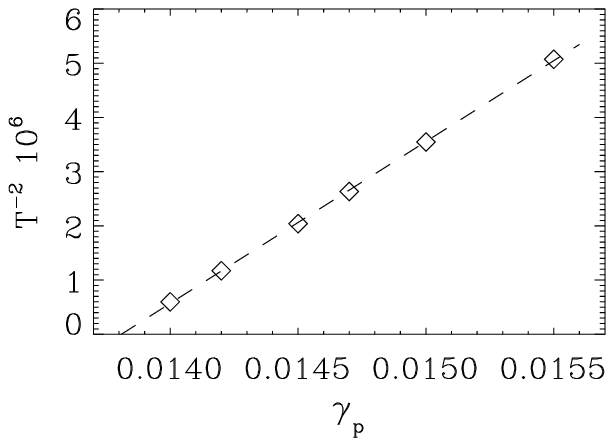}
\caption{Period of the oscillations ($T$) of $Q$ and $R$  obtained from the 
numerical solution of the coupled CGLE 
using the parameters given in Fig. 1.
The dashed line is a least squares fitting from which
$\gamma_{Pc} = 0.0138$.  
}
\end{figure}
\begin{figure} 
\vskip 5.5 cm 
\includegraphics{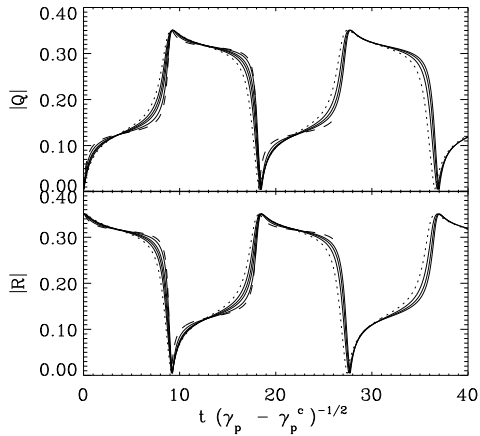}
\caption{The amplitudes of $Q$ and $R$ of the solution shown in
Fig. 2 exhibit the periodic oscillations given by Eq. (13).
The time has been scaled using the value of $\gamma_{Pc}$ obtained from
Fig. 3. The dotted line corresponds to $\gamma_p=0.0155$, the dashed
line to $\gamma_p=0.0140$, and the lines in between to the other points
of Fig. 3.}
\end{figure}

Close to the instability at $\gamma_{Pc}$,
the time dependent behavior of the solution can be obtained
reducing
the problem to a phase dynamics by elimination of the variable $X$.
We have (in the limit $\beta = 0$),
 \begin{eqnarray}
  \dot \Phi &=&  - { (\mu - k^2)(1- \gamma )
 \sin4\Phi \over 5+3\gamma -  (1- \gamma ) 
 \cos4\Phi } +\gamma_P\>,
  \end{eqnarray}
which for $\gamma_P \ge \gamma_{Pc}$ yields the following time behavior:
\begin{eqnarray} 
\tan (2 \Phi) &=& \tan (2 \Phi_c)
\left(1 +
\sqrt{ 2 (\gamma_P - \gamma_{Pc}) / \gamma_{Pc}  } \right.
\nonumber\\
 && \left.
\tan \left(
(5+3 \gamma)
\sqrt{ \gamma_{Pc} }
\sqrt{ \frac{\gamma_P - \gamma_{Pc}}{(1+\gamma)(3+\gamma)} }
\> t
\right)
\right)\>.
\label{oscillation}
\end{eqnarray}
where, $\Phi_c=\Phi_o$, Eq. (\ref{Phi}), for $\gamma_P=\gamma_{Pc}$.

An approximative analysis of Eqs. (\ref{polares}) for $\beta \ne 0$ 
indicates that this parameter appears squared in the expressions for
$\gamma_{Pc}$, $X$ and $\Phi_c$. Therefore, for small $\beta$, the 
previous analysis is still meaningful as explicitly seen in the numerical 
results of 
Figs. 3 and 4.

In summary, in the absence of forcing, and for the parameter regime 
considered here, there are 
solutions for the amplitudes $A$ and $B$,
of the coupled CGLE which correspond to linearly polarized standing waves. 
We have shown that an imaginary coupling between them
transforms these solutions into standing waves with spatially periodic 
elliptic polarization. Increasing the forcing, an instability of these
solutions, via the unlocking of the underlying wave amplitudes,
appears and the solutions acquire a time-periodic behavior. Locally,
this bifurcation is of the saddle-node type, but the presence 
of heteroclinic connections between the fixed points gives rise to the 
appearance of a limit cycle when stable and unstable points merge.

\vskip 0.2 cm

Financial support from 
DGICYT Projects PB94-1167 and PB94-1172 is acknowledged.

\end{twocolumns}

\vskip -0.2 cm 
\begin{references}
%
\bibitem[+]{Daniel} Director of Research at the Belgian National Fund for 
Scientific Research. Center for Nonlinear Phenomena and Complex 
Systems, Universit\'e Libre de Bruxelles, Campus Plaine, Blv. du Triomphe 
B.P 231, 1050 Bruxelles.
\bibitem{CrossHoh} M.C. Cross and P.C. Hohenberg, 
Rev. Mod. Phys. {\bf 65} , 854 (1993).
\bibitem{NewMol} A.C. Newell and J.V. Moloney, {\it Nonlinear Optics} 
(Addison-Wesley, 
Redwood City, 1992 ).
\bibitem{gollub} M. Lowe and J.P. Gollub,  
Phys. Rev., {\bf A31},  3893 (1985).
\bibitem{DW} D. Walgraef,  Europhys.Lett.,  {\bf 7},  
485 (1988). 
\bibitem{HRiecke} H.Riecke, J.D.Crawford and E.Knobloch,  
Phys. Rev. Lett. , {\bf 61},  1942 (1988).
\bibitem{Dangel} G. Dangelmayr and E. Knobloch, Nonlinearity {\bf 4}, 399 
(1991). 
\bibitem{tredicce} E.J. D'Angelo, E. Izaguirre, G.B. Mindlin, G. Huyet, L. Gil, 
and J.R. Tredicce, Phys. Rev. Lett. {\bf 68}, 3702 (1992).  
\bibitem{Douady} S. Douady, S. Fauve and O. Thual,
Europhysics Letters, {\bf 10}, 309 (1989).
\bibitem{kolodner} P. Kolodner, Phys. Rev. A {\bf 44}, 6448 (1991); 
Phys. Rev. Lett. 
{\bf 66}, 1165 (1991). 
\bibitem{liqcrys} S. Kai ed., Physics of Pattern Formation in Complex
Dissipative Systems, World Scientific, Singapour (1992).
\bibitem{valpo} J. Mart\'\i nez-Mardones, R. Tiemann, W. Zeller, and C. 
P\'erez-Garc\'\i a, Int. J. Bifurc. Chaos {\bf 4}, 5 (1994). 
\bibitem{singh} S. Singh, Phys. Rep. {\bf 108}, 217 (1984).
\bibitem{etrich} C. Etrich, P. Mandel, R. Neelen, R.J.C. Spreeuw, J.P. 
Woerdman, Phys. Rev. A {\bf 46}, 525 (1992).
\bibitem{maxi} M. San Miguel, Phys. Rev. Lett. {\bf 75}, 425 (1995).
\bibitem{Andronov} A.A. Andronov, A.A. Vitt and S.E. Khaikin, 
{\sl Theory of oscillators}, Pergamon Press, Oxford (1966).
\bibitem{excitable} S.C. Mueller, P. Coullet and D. Walgraef, {\sc chaos}
{\bf 4}, 3, 439 (1994).
\bibitem{meron} E. Meron, Phys. Rep. {\bf 218}, 1 (1992). 

  
\end{references}
\end{document}